\documentclass[10pt,twocolumn,letterpaper]{article}

\usepackage[pagenumbers]{cvpr} 

\usepackage{multirow}
\usepackage{bbding}
\usepackage{pifont}

\usepackage[dvipsnames]{xcolor}
\newcommand{\red}[1]{{\color{red}#1}}
\newcommand{\green}[1]{{\color[rgb]{0.145,0.643,0.105}#1}}
\newcommand{\yellow}[1]{{\color[rgb]{0.85,0.85,0.22}#1}}
\newcommand{\blue}[1]{{\color[rgb]{0.1,0.1,0.95}#1}}

\definecolor{cvprblue}{rgb}{0.21,0.49,0.74}
\usepackage[pagebackref,breaklinks,colorlinks,citecolor=cvprblue]{hyperref}

\def\paperID{1661} 
\def\confName{CVPR}
\def\confYear{2024}

\title{C$^\text{2}$RV: Cross-Regional and Cross-View Learning for Sparse-View \\ CBCT Reconstruction}

\author{
\vspace{2pt}
Yiqun Lin$^{1}$ \quad 
Jiewen Yang$^{1}$ \quad 
Hualiang Wang$^{1}$ \quad 
Xinpeng Ding$^{1}$ \quad 
Wei Zhao$^{2}$\thanks{Corresponding Authors} \quad 
Xiaomeng Li$^{1*}$
\\
$^1$The Hong Kong University of Science and Technology \quad
$^2$Beihang University 
\\
{\tt\small
\{ylindw, jyangcu, hwangfd, xdingaf\}@connect.ust.hk,}
\vspace{-1pt}
\\
{\tt\small 
zhaow20@buaa.edu.cn, 
eexmli@ust.hk
} 
}

\newcommand{\tabvs}{-7pt}
\newcommand{\figvs}{-0.65cm}

\newcommand{\nickname}{C$^\text{2}$RV}

\begin{document}
\maketitle
\begin{abstract}

Cone beam computed tomography (CBCT) is an important imaging technology widely used in medical scenarios, such as diagnosis and preoperative planning. Using fewer projection views to reconstruct CT, also known as sparse-view reconstruction, can reduce ionizing radiation and further benefit interventional radiology. 
Compared with sparse-view reconstruction for traditional parallel/fan-beam CT, CBCT reconstruction is more challenging due to the increased dimensionality caused by the measurement process based on cone-shaped X-ray beams. As a 2D-to-3D reconstruction problem, although implicit neural representations have been introduced to enable efficient training, only local features are considered and different views are processed equally in previous works, resulting in spatial inconsistency and poor performance on complicated anatomies. 
To this end, we propose \nickname{} by leveraging explicit multi-scale volumetric representations to enable cross-regional learning in the 3D space. Additionally, the scale-view cross-attention module is introduced to adaptively aggregate multi-scale and multi-view features. Extensive experiments demonstrate that our \nickname{} achieves consistent and significant improvement over previous state-of-the-art methods on datasets with diverse anatomy. 
Code is available at \href{https://github.com/xmed-lab/C2RV-CBCT}{\tt\small https://github.com/xmed-lab/C2RV-CBCT}.

\end{abstract}

\section{Introduction} \label{sec:intro}

\begin{figure}[t]
\centering 
\includegraphics[width=1.0\linewidth]{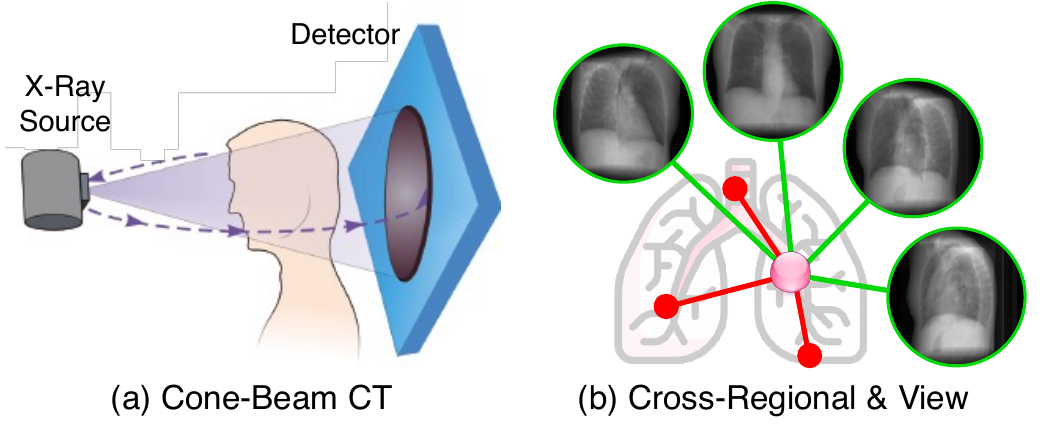}
\vspace{-0.7cm}
\caption{(a) Cone-shaped X-ray beams are emitted from the scanning source and a 2D array of detectors measures the transmitted radiation. (b) Cross-regional (\red{red}) and cross-view (\green{green}) feature learning to enhance point-wise representation.}
\label{fig:teaser}
\end{figure}

Computed tomography (CT) has become an indispensable technique used for medical diagnostics, providing accurate and non-invasive visualization of internal anatomical structures. Compared with conventional CT (fan/parallel-beam), cone-beam CT (CBCT) offers advantages, including faster acquisition and improved spatial resolution~\cite{scarfe2006clinical}. Typically, hundreds of projections are required to produce a high-quality CT scan involving high radiation doses from X-rays. However, high radiation dose exposure to patients can be a concern in clinical practice, limiting its use in scenarios like interventional radiology. Hence, reducing the number of projections can be one of the ways to reduce the radiation doses, which is also known as sparse-view reconstruction.

Over the past decades, there have been many research works studying the sparse-view problem for conventional CT by formulating the reconstruction as a mapping from 1D projections to a 2D CT slice, where generation-based techniques~\cite{jin2017deep, han2016deep, zhang2018sparse, wang2018conditional, huang2018metal, ma2023freeseed, he2020radon, wu2021drone, ma2023freeseed} are proposed to operate on the image or projection domains. However, the measurements of cone-beam CT are 2D projections (Figure~\ref{fig:teaser}a), resulting in increased dimensionality compared with conventional CT. This means that extending previous conventional CT reconstruction methods to CBCT will encounter issues~\cite{lin2023learning} such as high computational cost.

Recently, implicit neural representations (INRs) have been widely used in 3D reconstruction, including novel view synthesis and object reconstruction. To handle sparse-view or even single-view scenarios, geometric priors (\eg, surface points~\cite{xu2022point} and normals~\cite{yin2022coordinates}) or parametric shape models~\cite{huang2020arch, zheng2021pamir, xiu2022icon, xiu2023econ} (\eg, SMPL~\cite{loper2015smpl} and SMPL-X~\cite{pavlakos2019expressive}) are incorporated to improve the robustness and generalization ability. 
However, unlike visible light, X-rays have a higher frequency and pass through the surfaces of many materials, hence, no depth or surface information can be measured in the projection. Additionally, it is difficult to build a CT-specific parametric model as the internal anatomies of the human body are more complicated than surface models. 

\begin{figure}[t]
\centering 
\includegraphics[width=1.0\linewidth]{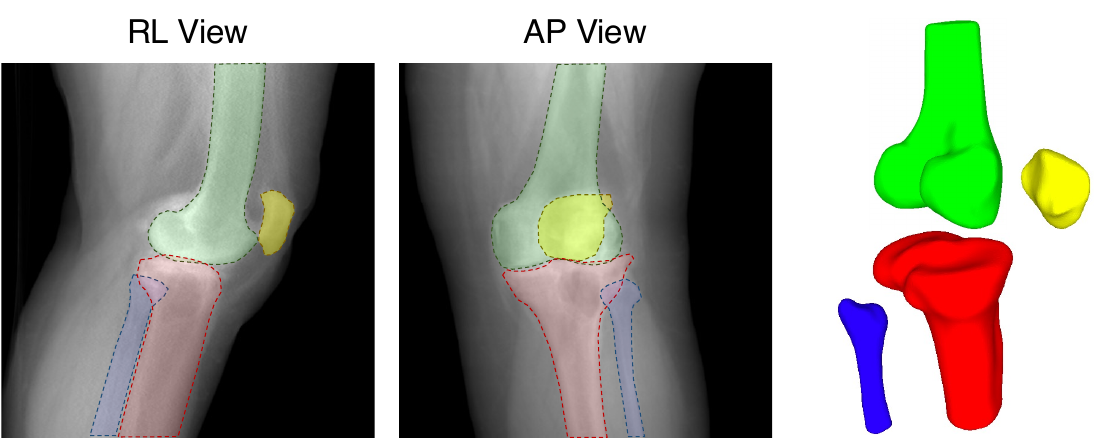}
\vspace{\figvs{}}
\caption{Right-left (RL) and anterior-posterior (AP) views of the knee. \green{Green}: femur. \red{Red}: tibia. \yellow{Yellow}: patella. \blue{Blue}: fibula. The patella and femur overlap in the AP view but not in the RL view.}
\label{fig:knee}
\end{figure}

Although INRs have been introduced to CBCT reconstruction in recent years, tens of views (\ie, 20-50) are still required for self-supervised NeRF-based methods~\cite{zha2022naf, fang2022snaf, shen2022nerp} due to the lack of prior knowledge. On the other hand, current data-driven methods like DIF-Net~\cite{lin2023learning} may suffer from poor performance when the anatomy has complicated structures for two possible reasons: 1.) local features queried from projections can be difficult to identify different organs that have low contrast in the projection; 2.) projections of different views are processed equally, while some views indeed present more information of specific organs than other views. For example, the right-left view shows the patella clearly, while it overlaps the femur in the anterior-posterior view; see Figure~\ref{fig:knee}.

To address the limitations of previous works, we propose a novel sparse-view CBCT reconstruction framework \nickname{} by leveraging cross-regional and cross-view feature learning to enhance point-wise representation (Figure~\ref{fig:teaser}b).
To be more specific, we first introduce multi-scale 3D volumetric representations (MS-3DV), where features are obtained by back-projecting multi-view features at different scales to the 3D space. 
Explicit MS-3DV enables cross-regional learning in 3D space, providing richer information that helps better identify different organs.
Hence, the feature of a point can be queried in a hybrid way, \ie, multi-scale voxel-aligned features from MS-3DV and multi-view pixel-aligned features from projections. Instead of considering queried features equally, scale-view cross-attention (SVC-Att) is then proposed to adaptively learn aggregation weights by self-attention and cross-attention. Finally, multi-scale and multi-view features are aggregated to estimate the attenuation coefficient.
We evaluate \nickname{} quantitatively and qualitatively on two CT datasets (\ie, chest and knee). Extensive experiments demonstrate that our proposed \nickname{} consistently outperforms previous state-of-the-art methods by a considerable margin under different experimental settings. 

The main contributions of this work are summarized as:

\begin{itemize}[itemsep=4pt,leftmargin=22pt]
    \item Multi-scale 3D volumetric representations (MS-3DV) to enable cross-regional learning in the 3D space;
    
    \item Scale-view cross-attention (SVC-Att) to adaptively aggregate multi-scale and multi-view features;

    \item \nickname{}, a novel sparse-view CBCT reconstruction framework, achieving state-of-the-art performance on datasets with diverse anatomy.

    \item Ablative studies to analyze the effectiveness and robustness of the proposed \nickname{}.
\end{itemize}


\section{Related Work}

In computer vision, especially 3D vision, the reconstruction problem has gained significant attention in recent years. In this section, we mainly review related work of sparse-view reconstruction on traditional parallel/fan-beam CT, cone-beam CT, and general 3D.

\subsection{Sparse-View CT Reconstruction}

Traditional parallel/fan-beam CT reconstruction can be regarded as reconstructing a 2D CT slice from 1D projections. Existing learning-based methods mainly include image-domain, projection-domain, and dual-domain methods. 
Specifically, image-domain methods~\cite{jin2017deep, han2016deep, zhang2018sparse, wang2018conditional, huang2018metal, ma2023freeseed} apply filtered back projection (FBP) to reconstruct a coarse CT slice with streak artifacts and utilize CNNs, such as U-Net~\cite{ronneberger2015u} and DenseNet~\cite{huang2017densely}, to denoise and refine details. 
When extending these methods to CBCT reconstruction, the network should be modified to 3D CNNs, resulting in a substantial increase in computational cost. Another way is to adopt these methods for slice-wise (2D) denoising~\cite{lahiri2022sparse}, while the 3D spatial consistency cannot be guaranteed.

Projection-domain methods directly operate on sparse-view 1D projections by mapping the projections to the CT slice~\cite{he2020radon} or recovering the full-view projections~\cite{wu2021drone}. Additionally, Song \etal~\cite{song2021solving} utilize score-based generative models and propose a sampling method to reconstruct an image consistent with both the measurement process and the observed measurements (\ie, projections). Chung \etal~\cite{chung2023solving} further incorporate 2D diffusion models into iterative reconstruction.
Dual-domain methods operate on both projection and image domains by combining the denoising processes of two domains~\cite{lin2019dudonet, ma2023freeseed} or modeling dual-domain consistency~\cite{wang2021dudotrans}.
However, projection-based operations cannot be extended to CBCT reconstruction as the measurement processes (cone-beam \vs parallel/fan-beam) are different.

\subsection{Sparse-View CBCT Reconstruction} \label{sec:rw_cbct}

Different from traditional parallel/fan-beam CT, the measurement of cone-beam CT is a 2D projection, which means the reconstruction should be formulated as reconstructing a 3D CT volume from multiple 2D projections. Conventional filtered back-projection (FDK~\cite{feldkamp1984practical}) and ART-based iterative methods~\cite{gordon1970algebraic, andersen1984simultaneous, pan2006variable} often suffer from heavy streaking artifacts and poor image quality when the number of projections is dramatically decreased. Recently, learning-based approaches are proposed for single/orthogonal-view CBCT reconstruction~\cite{jiang2022mfct, shen2019patient, ying2019x2ct, kyung2023perspective}, while these methods are specially designed for single/orthogonal-view reconstruction~\cite{jiang2022mfct, ying2019x2ct, kyung2023perspective} or patient-specific data~\cite{shen2019patient}, making them difficult to extend to general sparse-view reconstruction.

On the other hand, implicit neural representations~\cite{mildenhall2021nerf, ruckert2022neat} have been introduced to represent CBCT as an attenuation~\cite{fang2022snaf, zha2022naf} or intensity~\cite{lin2023learning} field. Self-supervised methods, including NAF~\cite{zha2022naf} and NeRP~\cite{shen2022nerp}, simulate the measurement process and minimize the error between real and synthesized projections. However, these methods require a long time for per-sample optimization and are only suitable for the reconstruction from tens of views (\ie, 20-50) due to the lack of prior knowledge. DIF-Net~\cite{lin2023learning}, as a data-driven method, formulates the problem as learning a mapping from sparse projections to the intensity field. Nevertheless, DIF-Net regards different projections equally, and only local semantic features are queried for each sampled point, leading to limited reconstruction quality when processing anatomies with complicated structures (\eg, chest).

\subsection{Sparse-View 3D Reconstruction}

In 3D computer vision, implicit representations have been widely used in novel-view synthesis~\cite{mildenhall2021nerf, yu2021pixelnerf, yin2022coordinates, xu2022point} and object reconstruction~\cite{park2019deepsdf, saito2019pifu, huang2020arch, zheng2021pamir, xiu2022icon, xiu2023econ}. 
For novel view synthesis, to extend NeRF~\cite{mildenhall2021nerf} to sparse-view scenarios, geometric priors like surface points~\cite{xu2022point} and normals~\cite{yin2022coordinates} are incorporated to improve the generalization ability and efficiency. 
For object reconstruction, particularly digital human reconstruction, previous works~\cite{huang2020arch, zheng2021pamir, xiu2022icon, xiu2023econ} leverage explicit parametric SMPL(-X)~\cite{loper2015smpl, pavlakos2019expressive} models to constrain surface reconstruction and improve the robustness.
However, there is no available depth or surface information in the attenuation fields of CBCT since X-rays penetrate right through many common materials, such as flesh.
SMPL(-X) are 3D parametric shape models specially designed for the surface of the human body, while the internal anatomy structures are too complicated to design a CT-specific parametric model. Therefore, parametric shape models cannot be used in sparse-view CBCT reconstruction.
Furthermore, cross-view relationships are rarely considered in surface-based reconstruction since one or two views are more practical and often sufficient to learn the sparse field with the above-mentioned priors.

\section{Methodology}

In this section, we first revisit the problem formulation of sparse-view CBCT reconstruction and the baseline DIF-Net proposed in \cite{lin2023learning}. Then, we formally introduce \nickname{}, consisting of multi-scale 3D volumetric representations (MS-3DV) and the scale-view cross-attention (SVC-Att) for cross-regional and cross-view learning.

\subsection{Revisit DIF-Net~\cite{lin2023learning}}
\label{sec:dif-net}

We follow previous works \cite{lin2023learning, zha2022naf} to formulate the CT image as a continuous implicit function $g$:~$\mathbb{R}^3 \rightarrow \mathbb{R}$, which defines the attenuation coefficient (same as ``intensity'' in \cite{lin2023learning}) $v \in \mathbb{R}$ of a point $p \in \mathbb{R}^3$ in the 3D space, \ie, $v = g(p)$. Hence, given $N$-view projections $\mathcal{I} = \{I_1, \dots, I_N\} \subset \mathbb{R}^{W\times H}$ ($W$ and $H$ are width and height) with known scanning parameters (\eg, viewing angles, distance of source to origin) during the measurement process, the reconstruction problem is formulated as a conditioned implicit function $g(\cdot)$ such that $v = g(\mathcal{I}, p)$. 

In practice, a 2D encoder-decoder (shared across different views) is used to extract multi-view feature maps $\mathcal{F} = \{\mathcal{F}_1, \dots, \mathcal{F}_N\} \subset \mathbb{R}^{C\times(W\times H)}$ from $N$-view projections $\mathcal{I}$, where $C$ is the output channel size of the decoder. For $i^\text{th}$ view, denote the projection function as $\pi_i$:~$\mathbb{R}^3 \rightarrow \mathbb{R}^2$, which maps a 3D point $p$ to the 2D plane where detectors are located such that $p'_i = \pi_i(p)$. Then, we define the view-specific pixel-aligned features of $p$ in $i^\text{th}$ view as 
\begin{equation}
\label{eq:interp}
\begin{split}
    \mathcal{F}_i(p) &= \text{Interp}\big(\mathcal{F}_i, \pi_i(p)\big) \\
                     &= \text{Interp}(\mathcal{F}_i, p'),
\end{split}
\end{equation}
where $\text{Interp}$:~$(\mathbb{R}^{C \times (D_1 \times \cdots \times D_k)}, \mathbb{R}^k) \rightarrow \mathbb{R}^C$ is $k$-linear interpolation. Particularly, $k =2 $ and $\text{Interp}(\cdot)$ is bilinear interpolation in the above equation.

Denoting multi-view pixel-aligned features of $p$ as $\mathcal{F}(p) = \big\{\mathcal{F}_1(p), \dots, \mathcal{F}_N(p)\big\} \subset \mathbb{R}^C$, the attenuation coefficient of $p$ is 
\begin{equation}
\label{eq:aggregate}
    v = g(\mathcal{I}, p) = \sigma\big(\mathcal{F}(p)\big),    
\end{equation}
where $\sigma(\cdot)$ is the aggregation function implemented with MLPs (or Max-Pooling + MLPs) in DIF-Net~\cite{lin2023learning}. Although the above formulation and implementation enable efficient training for high-resolution sparse-view reconstruction, only local pixel-aligned features queried from projections are considered and different views are processed equally, leading to poor performance on complicated anatomies; see analysis in Sec.~\ref{sec:intro}~\&~\ref{sec:rw_cbct} and results in Table~\ref{tab:all}. To this end, we propose \nickname{} and will introduce it in detail in the following section.

\begin{figure*}[t]
\centering 
\includegraphics[width=1.0\textwidth]{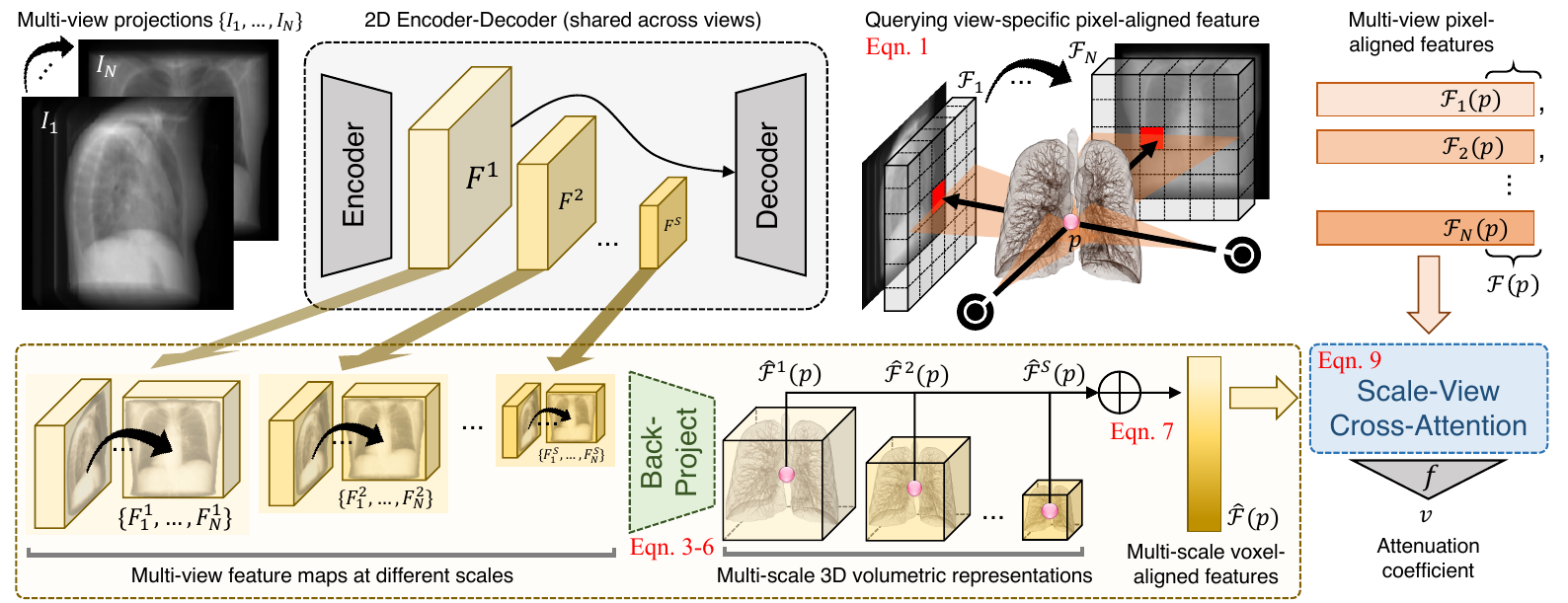}
\vspace{\figvs{}}
\caption{The overview of the proposed sparse-view reconstruction framework \nickname{}. 
Given multi-view projections, a 2D encoder-decoder is applied to extract view-wise feature map $\mathcal{F}_i$ for querying the pixel-aligned feature $\mathcal{F}_i(p)$. 
Additionally, the output feature map $F^1$ of the encoder is downsampled to obtain multi-scale feature maps. At each scale $s$, multi-view features are back-projected to the 3D space and gathered to form the 3D volumetric representation $\hat{\mathcal{F}}^s$ for querying the voxel-aligned feature $\hat{\mathcal{F}}^s(p)$.
Finally, multi-scale voxel-aligned features and multi-view pixel-aligned features are aggregated via scale-view cross-attention modules to estimate the attenuation coefficient.}
\label{fig:overview}
\end{figure*}

\subsection{\nickname{} Framework}

\nickname{} (\underline{\textbf{C}}ross-\underline{\textbf{R}}egional and \underline{\textbf{C}}ross-\underline{\textbf{V}}iew Learning) framework is developed based on DIF-Net~\cite{lin2023learning} to address the above-mentioned limitations. The framework overview is shown in Figure~\ref{fig:overview}. Specifically, multi-scale 3D volumetric representations (MS-3DV) are obtained by back-projecting multi-view feature maps at different scales to the 3D space. Hence, multi-scale voxel-aligned features and multi-view pixel-aligned features are adaptively aggregated by scale-view cross-attention (SVC-Att) modules to estimate the attenuation coefficient.

\vspace{6pt}
\noindent
\textbf{Low-Resolution 3D Volumetric Representation.} 
A 3D volumetric space $\mathcal{S} \in \mathbb{R}^{3\times(r\times r\times r)}$ is defined by voxelizing the 3D space with a low resolution $r \leq 16$. Let $F_i \in \mathbb{R}^{c\times w \times h}$ be the intermediate feature map of the encoder-decoder given the projection of $i^\text{th}$ view. The volumetric feature space $\hat{F} \in \mathbb{R}^{c\times (r\times r\times r)}$ defined over $\mathcal{S}$ is produced by back-projecting multi-view feature maps into $\mathcal{S}$, \ie,
\begin{equation}
    \label{eq:back-proj}
    \hat{F} = \text{Back-Project}\big(\{F_1, \dots, F_N\}, \mathcal{S}\big),
\end{equation}
where the feature of a voxel $q$ in $\mathcal{S}$ is 
\begin{equation}
\begin{split}
    \hat{F}(q) &= \varphi\Big(\big\{F_1(q), \dots, F_N(q)\big\}\Big), \\
    \text{where~} F_i(q) &= \text{Interp}\big(F_i, \pi_i(q)\big),
\end{split}
\end{equation}
and $\varphi(\cdot)$ is the aggregation function, implemented with Max-Pooling in practice. Therefore, 3D convolutional layers (denoted as $\phi$) can be followed for efficient cross-regional feature learning, \ie,
\begin{equation}
    \label{eq:bp-conv}
    \hat{\mathcal{F}} = \phi(\hat{F}).
\end{equation}

\vspace{6pt}
\noindent
\textbf{MS-3DV: Multi-Scale 3D Volumetric Representations.} 
To further improve the robustness of reconstructing different anatomical structures, we propose to leverage multi-scale 3D volumetric representations. To be specific, given the projection of $i^\text{i}$ view, denote the output feature map of the encoder as $F^1_i$, then a sequence of downsampling operators $\rho$ are applied to produce multi-scale feature maps $\{F^1_i, \dots, F^S_i\}$, where $F^s_i = \rho_{s-1}(F^{s-1}_i)$ for $s \in \{2, \dots, S\}$, and $S$ is the total number of scales. 
Then, we define multi-scale 3D voxelized space $\{\mathcal{S}^1, \cdots, \mathcal{S}^S\}$ with different resolutions $\{r^1, \dots, r^S\}$, and back-project (Eqn.~\ref{eq:back-proj} and \ref{eq:bp-conv}) multi-view feature maps of each scale to obtain multi-scale 3D volumetric representations (MS-3DV) $\{\hat{\mathcal{F}}^1, \dots, \hat{\mathcal{F}}^S\}$, where
\begin{equation}
    \hat{\mathcal{F}}^s = \phi^s \Big(\text{Back-Project}\big(\{F_1^s, \dots, F_N^s\}, \mathcal{S}^s\big)\Big),
\end{equation}
for $s \in \{1, \dots, S\}$. Hence, in addition to multi-view pixel-aligned features directly queried from view-specific feature maps, we incorporate multi-scale voxel-aligned features for the point $p$ into the estimation of the attenuation coefficient, as given by
\begin{equation}
\label{eq:ms-3dv}
    \hat{\mathcal{F}}(p) = \text{MLPs}\Big(\text{Concat}\big[\hat{\mathcal{F}}^1(p), \dots, \hat{\mathcal{F}}^S(p)\big]\Big),
\end{equation}
where $\hat{\mathcal{F}}^s(p) = \text{Interp}(\hat{\mathcal{F}}^s, p)$, $\text{Concat}[\cdot]$ indicates concatenation, and multi-layer perceptrons (MLPs) map the channel size of concatenated voxel-aligned features to be consistent with pixel-aligned features (Eqn.~\ref{eq:interp}), \ie, $C$.

\begin{figure}[t]
\centering 
\includegraphics[width=1.0\linewidth]{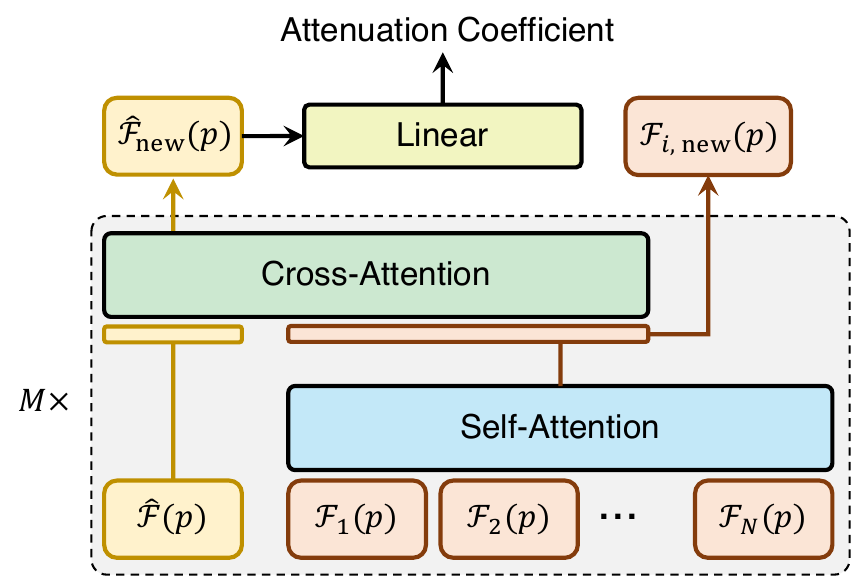}
\vspace{\figvs{}}
\caption{The overview of scale-view cross attention (SVC-Att) module. In each SVC-Att module, a self-attention is first applied to multi-view features, and then a cross-attention is followed to conduct attention between multi-scale features and multi-view features. $M$ SVC-Att modules are stacked and finally followed by a linear layer to estimate the attenuation coefficient.}
\label{fig:att}
\end{figure}

\vspace{6pt}
\noindent
\textbf{SVC-Att: Scale-View Cross-Attention.}
We first recall the definition of cross-attention (C-Att)~\cite{vaswani2017attention} given the reference features $F_r \in \mathbb{R}^{L_r \times C_r}$ and source features $F_s \in \mathbb{R}^{L_s \times C_s}$,
\begin{equation}
\begin{split}
    &\text{C-Att}(F_r, F_s) = \text{softmax}(\frac{QK^T}{\sqrt{C_d}}) V, \\
    \text{where}~&Q = F_r M_q,~K = F_s M_k,~V = F_s M_v,
\end{split}
\end{equation}
and $M_q \in \mathbb{R}^{C_r \times C_d}$, $M_k, M_v \in \mathbb{R}^{C_s \times C_d}$. Self-attention (S-Att) can be regarded as a special case of cross-attention where we let $F_s = F_r$.

For a point $p$, let $\mathcal{F}(p) = \big\{\mathcal{F}_1(p), \dots, \mathcal{F}_N(p)\big\} \subset \mathbb{R}^C$ denote multi-view pixel-aligned features queried from projections (Eqn.~\ref{eq:interp}), and $\hat{\mathcal{F}}(p) \in \mathbb{R}^C$ indicate the multi-scale voxel-aligned features queried from MS-3DV (Eqn.~\ref{eq:ms-3dv}). The scale-view cross-attention (SVC-Att) module is proposed to adaptively aggregate the above features. As shown in Figure~\ref{fig:att}, a self-attention module is first applied to conduct cross-view attention on multi-view features $\mathcal{F}(p)$. Then, a cross-attention attention module takes multi-scale features as the reference and the output of the self-attention module as the source to conduct attention between multi-scale and multi-view features. To formulate,
\begin{equation}
\label{eq:att}
\begin{split}
    \hat{\mathcal{F}}_\text{new}(p) &= \text{SVC-Att}\big(\hat{\mathcal{F}}(p), \mathcal{F}(p)\big) \\
    &= \text{C-Att}\big(\hat{\mathcal{F}}(p), \mathcal{F}_\text{new}(p)\big), \\
    \text{where~} \mathcal{F}_\text{new}(p) &= \text{S-Att}\big(\mathcal{F}(p)\big).
\end{split}
\end{equation}
In practice, $M$ SVC-Att modules are stacked and a linear layer is followed to estimate the attenuation coefficient.

\subsection{Network Training}

\begin{table*}[t]
{
\centering
\caption{Comparison\textsuperscript{1} of different methods on two CT datasets (\ie, chest and knee) with various numbers of projection views. The resolution of the reconstructed CT is 256$^\text{3}$. The reconstruction results are evaluated with PSNR (dB) and SSIM ($\times$10$^\text{-2}$), where higher PSNR/SSIM indicate better performance. The best values are \textbf{bolded} and the second-best values are \underline{underlined}.}
\label{tab:all}
\vspace{\tabvs{}}
\setlength{\tabcolsep}{6.5pt}
\resizebox{1.0\linewidth}{!}{
\begin{tabular}{l|c|ccc|ccc}
\toprule[1.2pt]
\multirow{2}{*}{Method} & \multirow{2}{*}{Type} & \multicolumn{3}{c|}{LUNA16~\cite{setio2017validation} (Chest CT)} & \multicolumn{3}{c}{Lin \etal~\cite{lin2023learning} (Knee CBCT)} \\ \cline{3-8}
 &  & 6-View & 8-View & 10-View & 6-View & 8-View & 10-View \\ 
 \hline \hline
FDK~\cite{feldkamp1984practical} & \multirow{4}{*}{\begin{tabular}[c]{@{}c@{}}Self-\\ Supervised\end{tabular}} & 15.29$|$27.80 & 16.54$|$28.05 & 17.36$|$29.06 & 18.42$|$30.56 & 19.83$|$32.42 & 20.95$|$34.55 \\
SART~\cite{andersen1984simultaneous} &  & 21.57$|$61.26 & 22.80$|$66.24 & 23.76$|$69.48 & 24.30$|$64.88 & 25.23$|$68.28 & 25.97$|$70.79 \\
NAF~\cite{zha2022naf} & & 18.76$|$39.02 & 20.51$|$46.09 & 22.17$|$52.57 & 20.11$|$47.35 & 22.42$|$55.19 & 24.26$|$61.72 \\
NeRP~\cite{shen2022nerp} & & 23.55$|$60.59 & 25.83$|$67.81 & 26.12$|$69.42 & 24.24$|$56.78 & 25.55$|$61.56 & 26.33$|$67.70 \\ 
\hline
FBPConvNet~\cite{jin2017deep} & \multirow{3}{*}{\begin{tabular}[c]{@{}c@{}}Data-Driven:\\ Denoising\end{tabular}} & 24.38$|$65.97 & 24.87$|$67.21 & 25.90$|$68.98 & 25.10$|$72.07 & 25.93$|$72.86 & 26.74$|$75.51 \\
FreeSeed~\cite{ma2023freeseed} & & \underline{25.59}$|$66.03 & \underline{26.86}$|$67.44 & \underline{27.23}$|$68.62 & 26.74$|$73.42 & 27.88$|$75.82 & 28.77$|$77.87 \\
BBDM~\cite{li2023bbdm} & & 24.78$|$65.80 & 25.81$|$67.06 & 26.35$|$68.71 & 26.58$|$74.42 & 28.01$|$75.71 & 28.90$|$77.26 \\
\hline
PixelNeRF~\cite{yu2021pixelnerf} & \multirow{3}{*}{\begin{tabular}[c]{@{}c@{}}Data-Driven:\\ INR-based\end{tabular}} & 24.66$|$66.49 & 25.04$|$68.24 & 25.39$|$70.62 & 26.10$|$79.96 & 26.84$|$81.33 & 27.36$|$82.49 \\
DIF-Net~\cite{lin2023learning} & & 25.55$|$\underline{73.19} & 26.09$|$\underline{76.96} & 26.69$|$\underline{78.56} & \underline{27.12}$|$\underline{80.74} & \underline{28.31}$|$\underline{82.03} & \underline{29.33}$|$\underline{84.98} \\
\nickname{} (\textit{ours}) & & \textbf{29.23}$|$\textbf{87.47} & \textbf{29.95}$|$\textbf{88.46} & \textbf{30.70}$|$\textbf{89.16} & \textbf{29.73}$|$\textbf{88.87} & \textbf{30.68}$|$\textbf{89.96} & \textbf{31.55}$|$\textbf{90.83} \\
\bottomrule[1.2pt]
\end{tabular}
}
}
{\small\textsuperscript{1}Compared with our camera-ready version, we have made the following updates: 1.) further tuned the hyperparameters for FDK and SART to improve their performance; 2.) re-evaluated SSIM by setting {\tt data\_range=1} in {\tt skimage.metrics.structural\_similarity} (in default, {\tt data\_range=2} for floating point inputs). For more detailed information about the evaluation, please visit our GitHub repository. The new results are also updated in subsequent tables.}
\end{table*}

We follow \cite{lin2023learning} to train the reconstruction network on a CT dataset, where the projections are simulated from the CT by digitally reconstructed radiographs (DRRs). 
Specifically, we denote the volumetric CT as $I_\text{ct} \in \mathbb{R}^{1\times (W_\text{ct}\times H_\text{ct}\times D_\text{ct})}$ and the projections as $\mathcal{I}$. Then, the ground-truth attenuation field defined over the continuous 3D space $\mathcal{P}$ is
\begin{equation}
    \mathcal{V} = \big\{v(p) = \text{Interp}(I_\text{ct}, p) ~\big|~ \forall p \in \mathcal{P}\big\},
\end{equation}
where $\text{Interp}(\cdot)$ is the interpolation operator (Eqn.~\ref{eq:interp}). The estimated attenuation field by \nickname{} is given as
\begin{equation}
    \hat{\mathcal{V}} = \big\{\hat{v}(p) = g(\mathcal{I}, p) ~\big|~ \forall p \in \mathcal{P}\big\}.
\end{equation}
Hence, the mean square error (MSE) as the objective function is used to compute point-wise estimation error,
\begin{equation}
\label{eq:loss}
    \mathcal{L}_\text{MSE}\big(\mathcal{V}, \hat{\mathcal{V}}\big) = \frac{1}{|\mathcal{P}|}\sum_{p \in \mathcal{P}} \Big(v(p) - \hat{v}(p)\Big)^2.
\end{equation}

During each training iteration, we randomly sample 10,000 points from $\mathcal{P}$ for loss calculation (Eqn.~\ref{eq:loss}) to reduce the memory requirements for efficient network optimization. During the inference, the 3D space is voxelized with a specified resolution (\eg, $256^3$), where the attenuation coefficient of a voxel is defined as the estimated attenuation coefficient of its centroid point by \nickname{}. This means that the resolution can be chosen based on the desired trade-off between image quality and reconstruction speed.

\vspace{3pt}
\noindent
\textbf{Implementation.}
In practice, we empirically choose $S = 3$, $r^1 = 16$, and $r^s = \frac{1}{2}r^{s-1}$ for $s \geq 2$.
We follow \cite{lin2023learning} to use U-Net~\cite{ronneberger2015u} with $C = 128$ output feature channels as the 2D encoder-decoder, where the size of encoder output $F^1$ is $\frac{W}{16} \times \frac{H}{16}$. $\phi(\cdot)$ in Eqn.~\ref{eq:bp-conv} is implemented with 3-layer 3D residual convolution that maps the channel size of $\hat{F}$ to $C$. 
For the aggregation method, $M=3$ SVC-Att modules are stacked, and attention modules are implemented as multi-head attention with 8 heads.
During training, the learnable parameters of \nickname{} are optimized using stochastic gradient
descent (SGD) with a momentum of 0.98 and an initial learning rate of 0.01. We train \nickname{} with 400 epochs and a batch size of 4. The learning rate is decreased by a factor of $(10^{-3})^{1/400}$ per epoch.

\section{Experiments}

To validate the effectiveness of our proposed \nickname{}, we conduct experiments on two CT datasets with different anatomies, including chest and knee. In addition to quantitative and qualitative evaluation, automatic segmentation is applied to sparse-view reconstruction results, showing the practical potential of reconstructed CT by \nickname{} in downstream applications.

\begin{figure*}[t]
\centering 
\includegraphics[width=1.0\textwidth]{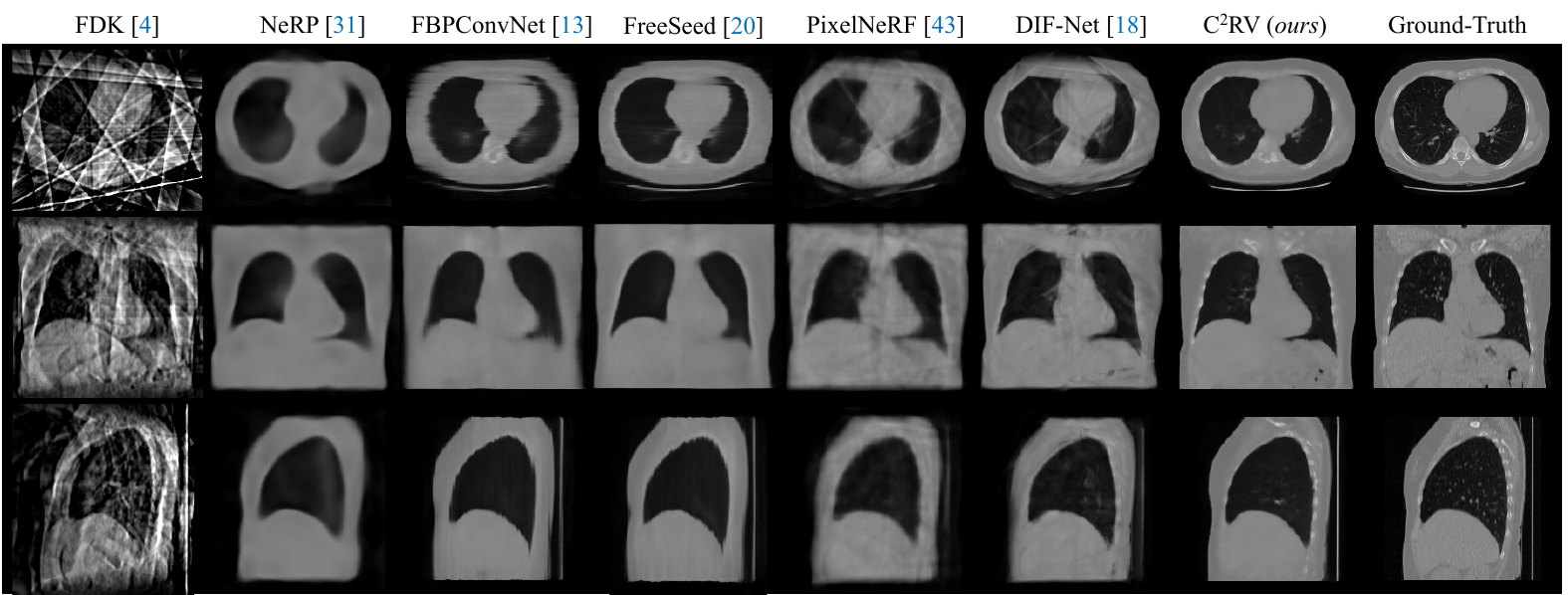}
\vspace{\figvs{}}
\caption{Visualization of 6-view reconstructed chest CT. From top to bottom: axial, coronal, and sagittal slices.}
\label{fig:vis_lung}
\end{figure*}

\subsection{Experimental Setting}

\noindent
\textbf{Dataset.} Experiments are conducted on two CT datasets, including a public chest CT dataset (LUNA16~\cite{setio2017validation}) and a private knee CBCT dataset collected by Lin~\etal~\cite{lin2023learning} (additional experiments on a dental CBCT dataset are provided in the supplementary). Specifically, LUNA16~\cite{setio2017validation} is composed of 888 chest CT scans with resolution ranging from 145$\times$145$\times$108 to 375$\times$375$\times$509~mm$^\text{3}$, split into 738 for training, 50 for validation, and 100 for testing; the knee dataset~\cite{lin2023learning} contains 614 knee CBCT scans with resolutions ranging from 236$\times$236$\times$167 to 500$\times$500$\times$416~mm$^\text{3}$, split into 464 for training, 50 for validation, and 100 for testing. We follow the data preprocessing of \cite{lin2023learning} to resample and crop (or pad) each CT to have isotropic spacing (\ie, 1.6~mm for chest and 0.8~mm for knee) and size of 256$^\text{3}$. Multi-view 2D projections are simulated by DRRs with a resolution of 256$^\text{2}$, and the viewing angles are uniformly selected in the range of 180$^\circ$ (half rotation).

\vspace{3pt}
\noindent
\textbf{Evaluation Metrics.} Following previous works~\cite{lin2023learning, zha2022naf, shen2022nerp}, two quantitative metrics, including peak signal-to-noise ratio (PSNR) and structural similarity (SSIM)~\cite{wang2004image}, are used to evaluate the reconstruction performance, where higher values indicate superior image quality.

\vfill

\subsection{Results}

\noindent
\textbf{Quantitative Evaluation.} 
We compare our \nickname{} with self-supervised methods, including FDK~\cite{feldkamp1984practical}, SART~\cite{andersen1984simultaneous}, NAF~\cite{zha2022naf}, and NeRP~\cite{shen2022nerp}, without requiring additional training data. 
We also compare data-driven approaches, including 2D denoising-based (\ie, FBPConvNet~\cite{jin2017deep}, FreeSeed~\cite{ma2023freeseed}, and BBDM~\cite{li2023bbdm}) and implicit neural representation (INR)-based (\ie, PixelNeRF~\cite{yu2021pixelnerf} and DIF-Net~\cite{lin2023learning}) methods. 
We conduct experiments with different numbers of projection views (\ie, 6-10) and the reconstruction resolution is 256$^\text{3}$. The results are shown in Table~\ref{tab:all}. 
Although DIF-Net~\cite{lin2023learning} can achieve satisfactory performance on knee CT, the performance drops dramatically when adapting to more complicated anatomical structures (\eg, chest), while our \nickname{} consistently performs well on different datasets. Additionally, when reconstructing from 6, 8, and 10 views, our \nickname{} outperforms previous state-of-the-art by a remarkable margin, \ie, 3.6/14.3, 3.1/11.5, and 3.5/10.6 PSNR/SSIM (dB/$\times$10$^\text{-2}$) on chest CT; and 2.6/8.1, 2.4/7.9, and 2.2/5.9 on knee CT. More importantly, Even with only 6 views, \nickname{} can reconstruct CT of better quality than other methods with 4 more views (\ie, 10 views).

\begin{figure}[t]
\centering 
\includegraphics[width=1.0\linewidth]{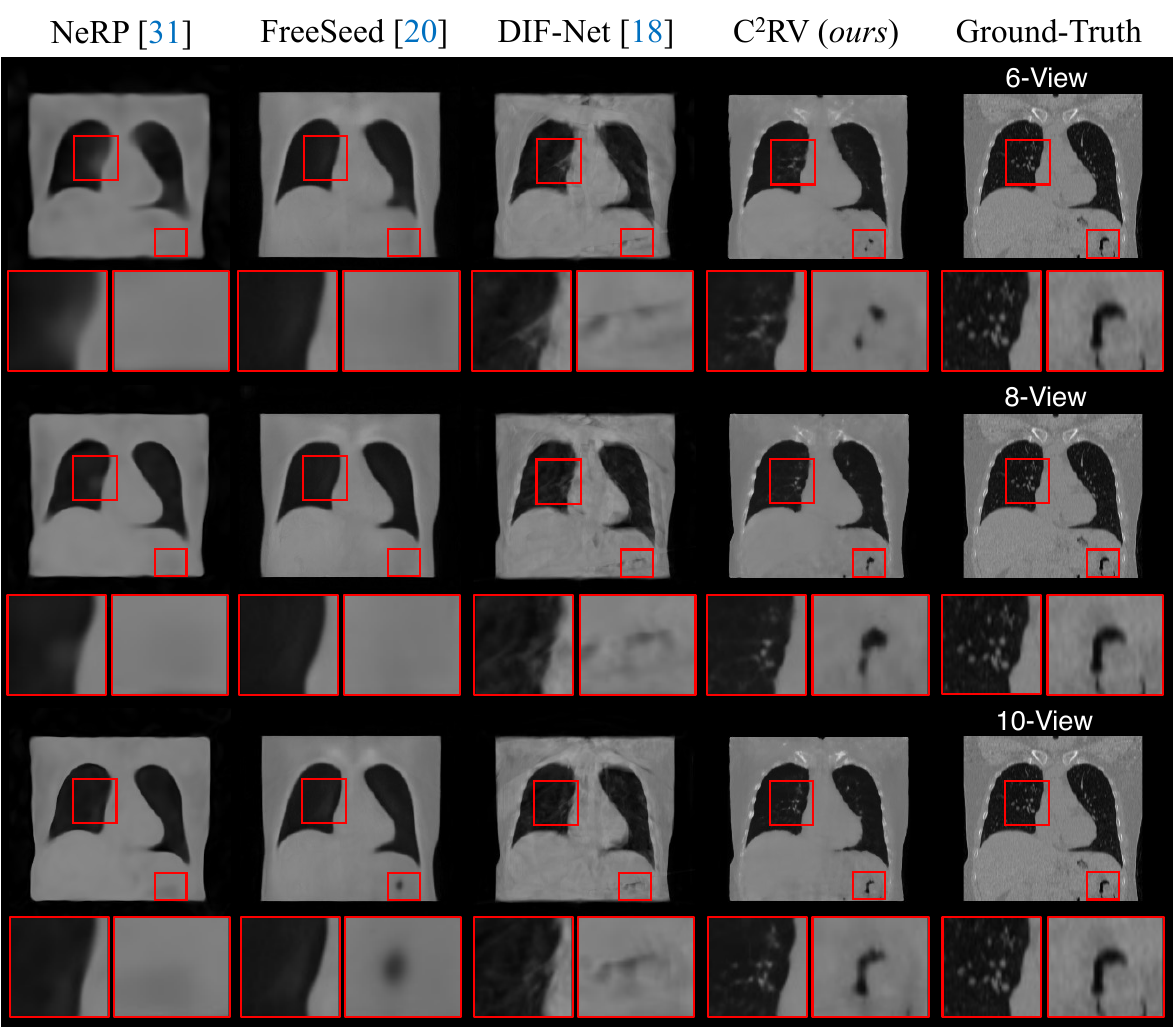}
\vspace{\figvs{}}
\caption{Visualization of examples reconstructed from different numbers of projection views, \ie, 6, 8, and 10. The highlighted regions (\red{red}) are zoomed in, showing richer details in our reconstructed results than in other methods.}
\label{fig:multi-view}
\end{figure}

\vspace{3pt}
\noindent
\textbf{Visual Comparison.} 
Examples of 6-view reconstruction are visualized in Figure~\ref{fig:vis_lung} for qualitative comparison. 
Due to the lack of sufficient projection views, reconstruction results of FDK~\cite{feldkamp1984practical} are full of streaking artifacts, and NeRP~\cite{shen2022nerp} can only reconstruct satisfactory contours of the body and lung. 
For FBPConvNet~\cite{jin2017deep} and FreeSeed~\cite{ma2023freeseed}, jitters appear near the boundary of the body and lung since they are 2D methods that reconstruct CT slice by slice. 
For PixelNeRF~\cite{yu2021pixelnerf} and DIF-Net~\cite{lin2023learning}, although the details are reconstructed better than others, there are still a few streaking artifacts and unclear contours. 
The reconstructed results of \nickname{} have clearer shape contours, better internal details, and almost no streaking artifacts.
Furthermore, Figure~\ref{fig:multi-view} shows the visualization of results reconstructed from different numbers of projection views, demonstrating a consistent conclusion with the above.

\begin{figure}[t]
\centering 
\includegraphics[width=1.0\linewidth]{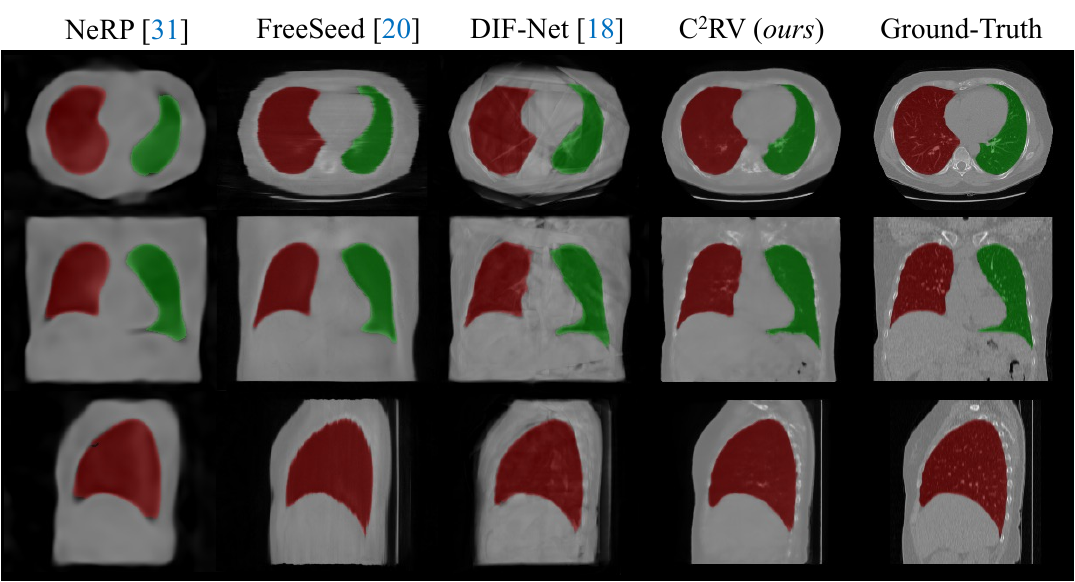}
\vspace{\figvs{}}
\caption{Visualization of lung segmentation on 6-view reconstructed chest CT. \red{Red}: left lung. \green{Green}: right lung.}
\label{fig:vis_seg}
\end{figure}

\vspace{3pt}
\noindent
\textbf{Downstream Evaluation.}
In addition to quantitative and qualitative evaluation, we validate the reconstructed CT on the downstream task, \ie, segmentation. Specifically, we utilize LungMask toolkit~\cite{hofmanninger2020automatic} to conduct left/right-lung segmentation on CT reconstructed by different methods.
As the results are shown in Table~\ref{tab:seg} and Figure~\ref{fig:vis_seg}, compared with other methods, the segmentation masks on the reconstructed CT of \nickname{} are more consistent with the segmentation on the ground-truth CT. This means our proposed \nickname{} has the potential to reconstruct high-quality CT that can be further applied in downstream scenarios.

\begin{table}[t]
\centering
\caption{Lung segmentation of 6-view reconstructed chest CT. Dice coefficient (\%, higher is better) and average surface distance (ASD, mm, lower is better) are evaluated. The best values are \textbf{bolded} and the second-best values are \underline{underlined}.}
\label{tab:seg}
\vspace{\tabvs{}}
\setlength{\tabcolsep}{3pt}
\resizebox{1.0\linewidth}{!}{
\begin{tabular}{l|cc|cc|cc}
\toprule[1.2pt]
\multirow{2}{*}{Method} & \multicolumn{2}{c|}{Recon.} & \multicolumn{2}{c|}{Left Lung} & \multicolumn{2}{c}{Right Lung} \\ \cline{2-7} 
 & PSNR & SSIM & Dice & ASD$\downarrow$ & Dice & ASD$\downarrow$ \\ \hline \hline
FDK~\cite{feldkamp1984practical} & 15.29 & 27.80 & 16.51 & 79.55 & 46.14 & 22.44 \\
NeRP~\cite{shen2022nerp} & 23.55 & 60.59 & 86.55 & 9.57 & 86.24 & 3.62 \\
FBPConvNet~\cite{jin2017deep} & 24.38 & 65.97 & 92.78 & 3.14 & 91.37 & 2.68 \\
FreeSeed~\cite{ma2023freeseed} & \underline{25.59} & 66.03 & \underline{95.16} & \underline{1.74} & 94.75 & \underline{1.77} \\
PixelNeRF~\cite{yu2021pixelnerf} & 24.66 & 66.49 & 91.00 & 5.31 & 91.66 & 3.67 \\
DIF-Net~\cite{lin2023learning} & 25.55 & \underline{73.19} & 94.45 & 2.51 & \underline{94.78} & 2.01 \\ \midrule[0.4pt]
\nickname{} (\textit{ours}) & \textbf{29.23} & \textbf{87.47} & \textbf{96.72} & \textbf{1.25} & \textbf{96.93} & \textbf{1.12} \\
\bottomrule[1.2pt]
\end{tabular}
}
\end{table}

\section{Ablation Study}

Ablation studies are conducted to explore the effectiveness of the proposed MS-3DV and SVC-Att, and different designs for MS-3DV. Moreover, we further analyze the robustness of our \nickname{} to varying viewing angles and noisy scanning parameters. All the following ablative experiments are conducted on 6-view reconstruction of chest CT with the resolution of 256$^\text{3}$.

\subsection{Proposed MS-3DV and SVC-Att}

\noindent
\textbf{Ablation on MS-3DV and SVC-Att.}
We regard DIF-Net~\cite{lin2023learning} as the baseline model and compare the reconstruction performance of introducing MS-3DV and SVC-Att. In DIF-Net, multi-view features are aggregated ($\sigma$ in Eqn.~\ref{eq:aggregate}) with MLPs or Max-Pooling + MLPs. Comparison is shown in Table~\ref{tab:aggregation}. In (+MS-3DV), multi-scale voxel-aligned features are concatenated with max-pooled multi-scale features. In (+SVC), we randomly initialize a learnable vector before training, as an alternative to the reference feature (\ie, $\hat{\mathcal{F}}(p)$ in Eqn.~\ref{eq:att}); also see Figure~\ref{fig:att}. Both MS-3DV and SVC-Att can improve the reconstruction performance, and the framework achieves new state-of-the-art performance by jointly incorporating the above two.

\vspace{3pt}
\noindent
\textbf{Different Designs for MS-3DV.} 
As shown in Table~\ref{tab:multi-scale}, we compare the performance of using different numbers of scales, and selections of initial feature map $F^1$ and resolution $r^1$. It is important to incorporate multi-scale features, which provide richer information than single-scale for identifying different anatomies, such as organs (\eg, lung) and bones (\eg, spine). We do not further increase the number of scales (\eg, 4) since the size of the feature map at the third scale is too small (\ie, 4$\times$4). For the choice of $F^1$, the output of the encoder is better as it contains more high-level features than the decoder. Empirically, the initial resolution of 16 is the best choice for the trade-off between the global (high-level) and local (details) features.

\subsection{Robustness Analysis}

Let $\mathcal{A} = \{\alpha_1, \dots, \alpha_N\}$ denote the viewing angles in the origional evaluation. The first experiment is conducted by choosing different viewing angles, \ie, $\mathcal{A}' = \{\alpha_i + \Delta\alpha~|~\alpha_i \in \mathcal{A}\}$, where $\Delta\alpha$ is the angle offset. As shown in Table~\ref{tab:robust}, the performance of \nickname{} is stable with varying angles. The second study is about the noisy scanning parameters. Taking the viewing angles as an example, we assume the measurement process is noisy, which means that multi-view projections are measured from $\mathcal{A}'=\{\alpha_i + \eta_i~|~\alpha_i \in \mathcal{A}\}$, where $\eta_i$ is the noise that obeys the uniform distribution $U(-\epsilon, +\epsilon)$. In this case, the projection function $\pi$ is still defined based on original viewing angles, \ie, $\mathcal{A}$, since the noise is unobservable. In Table~\ref{tab:robust}, we consider two scanning parameters, including the viewing angle, and the distance of source to origin, which are major factors related to the formulation of the projection function (see Appendix in \cite{lin2023learning}). Experiments show that our \nickname{} is robust to slight shifts in scanning parameters.

\vfill 

\begin{table}[t]
\caption{Ablation study on different aggregation methods (M.: MLPs~\cite{lin2023learning}, Max-M.: Max-Pooling + MLPs~\cite{lin2023learning}, SVC: our proposed scale-view cross-attention) and multi-scale 3D volumetric representations (MS-3DV). PSNR (dB) and SSIM ($\times$10$^\text{-2}$) are evaluated on 6-view reconstruction of chest CT.}
\label{tab:aggregation}
\vspace{\tabvs{}}
\setlength{\tabcolsep}{4pt}
\resizebox{1.0\linewidth}{!}{
\begin{tabular}{l|ccc|c|cc}
\toprule[1.2pt]
\multirow{2}{*}{Method} & \multicolumn{3}{c|}{Aggregation} & \multirow{2}{*}{MS-3DV} & \multirow{2}{*}{PSNR} & \multirow{2}{*}{SSIM} \\ \cline{2-4}
 & M. & Max-M. & SVC &  &  &  \\ \hline \hline
\multirow{2}{*}{DIF-Net~\cite{lin2023learning}} & \ding{51} &  &  &  & 25.55 & 73.19 \\
 &  & \ding{51} &  &  & 25.62 & 73.80 \\ \hline
~~~~+MS-3DV &  & \ding{51} &  & \ding{51} & 26.62 & 78.35 \\
~~~~+SVC &  &  & \ding{51} &  & 27.84 & 83.18 \\ \hline
\nickname{} (\textit{ours}) &  &  & \ding{51} & \ding{51} & \textbf{29.23} & \textbf{87.47} \\
\bottomrule[1.2pt]
\end{tabular}
}
\end{table}

\begin{table}[t]
\caption{Ablation study on the number of scales, the initial feature map $F^1$, and the initial resolution $r^1$. The selection of $F^1$ can be the final-layer feature map of the encoder or decoder. PSNR and SSIM are evaluated on 6-view reconstruction of chest CT.}
\label{tab:multi-scale}
\vspace{\tabvs{}}
\setlength{\tabcolsep}{6pt}
\resizebox{1.0\linewidth}{!}{
\begin{tabular}{ccc|ll}
\toprule[1.2pt]
\# Scales & $F^1$ & $r^1$ & PSNR (dB) & SSIM (10$^\text{-2}$) \\ \hline \hline
\red{\textbf{1}} & Encoder & 16 & 28.98 \footnotesize{\red{($-$0.25)}} & 87.06 \footnotesize{\red{($-$0.41)}} \\ 
\red{\textbf{2}} & Encoder & 16 & 29.09 \footnotesize{\red{($-$0.14)}} & 87.12  \footnotesize{\red{($-$0.35)}} \\ \hline
3 & \red{\textbf{Decoder}} & 16 & 28.57 \footnotesize{\red{($-$0.66)}} & 86.30 \footnotesize{\red{($-$1.17)}} \\ \hline
3 & Encoder & \red{\textbf{12}} & 28.96 \footnotesize{\red{($-$0.27)}} & 87.35 \footnotesize{\red{($-$0.12)}} \\
3 & Encoder & \red{\textbf{24}} & 29.23 \footnotesize{($-$0.00)} & 87.44 \footnotesize{\red{($-$0.03)}} \\ \hline
3 & Encoder & 16 & \textbf{29.23} & \textbf{87.47} \\
\bottomrule[1.2pt]
\end{tabular}
}
\end{table}

\begin{table}[]
\caption{Robustness analysis on varying angles and noisy scanning parameters, including viewing angles and the distance of source to origin (DSO). For noisy scanning parameters, the noisy offsets obey the uniform distribution, \ie, $U(-\epsilon, +\epsilon)$. PSNR and SSIM are evaluated on 6-view reconstruction of chest CT.}
\label{tab:robust}
\vspace{\tabvs{}}
\setlength{\tabcolsep}{6pt}
\resizebox{1.0\linewidth}{!}{
\begin{tabular}{c|cc|ll}
\toprule[1.2pt]
\multirow{2}{*}{\begin{tabular}[c]{@{}c@{}}Varying\\ Angles\end{tabular}} & \multicolumn{2}{c|}{Noisy Parameters} & \multirow{2}{*}{PSNR (dB)} & \multirow{2}{*}{SSIM (10$^\text{-2}$)} \\ \cline{2-3}
 & \multicolumn{1}{c|}{Angles} & DSO &  &  \\ \hline \hline
0$^\circ$ & \multicolumn{1}{c|}{-} & - & 29.23 & 87.47 \\ \hline
\red{\textbf{$+$10$^\circ$}} & \multicolumn{1}{c|}{\multirow{2}{*}{-}} & \multirow{3}{*}{-} & 29.24 \footnotesize{\green{($+$0.01)}} & 87.48 \footnotesize{\green{($+$0.01)}} \\
\red{\textbf{$+$20$^\circ$}} & \multicolumn{1}{c|}{} &  & 29.23 \footnotesize{($-$0.00)} & 87.47 \footnotesize{($-$0.00)} \\ \hline
\multirow{2}{*}{0$^\circ$} & \multicolumn{1}{c|}{\red{\textbf{$\pm$0.5$^\circ$}}} & \multirow{2}{*}{-} & 28.98 \footnotesize{\red{($-$0.25)}} & 87.12 \footnotesize{\red{($-$0.35)}} \\
 & \multicolumn{1}{c|}{\red{\textbf{$\pm$1.0$^\circ$}}} &  & 28.18 \footnotesize{\red{($-$1.05)}} & 86.03 \footnotesize{\red{($-$1.44)}} \\ \hline
\multirow{2}{*}{0$^\circ$} & \multicolumn{1}{c|}{\multirow{2}{*}{-}} & \red{\textbf{$\pm$2mm}} & 29.04 \footnotesize{\red{($-$0.19)}} & 87.24 \footnotesize{\red{($-$0.23)}} \\
 & \multicolumn{1}{c|}{} & \red{\textbf{$\pm$3mm}} & 27.85 \footnotesize{\red{($-$1.38)}} & 85.93 \footnotesize{\red{($-$1.54)}} \\
 \bottomrule[1.2pt]
\end{tabular}
}
\end{table}

\section{Conclusion}

In this work, we propose a novel framework, namely \nickname{}, for sparse-view cone-beam CT reconstruction. The novelties are mainly composed of 1.) multi-scale 3D volumetric representations (MS-3DV) to enable efficient cross-regional feature learning in the 3D space, and 2.) scale-view cross-attention (SVC-Att) to adaptively aggregate multi-scale and multi-view features.
Our \nickname{} shows superior reconstruction performance compared with previous state-of-the-art, the practical potential of reconstructed CT in downstream applications, and robustness to slightly noisy measurement processes.
Although our \nickname{} performs well in a specific dataset, it will fail when adapting to other datasets with unseen anatomies (\eg, chest$\rightarrow$head) as \nickname{} only learns the dataset-specific distribution priors. Hence, it would also be important to improve the few-shot or even zero-shot adaptation ability by introducing new training schemes or network frameworks, which will be left as our future works.

\vspace{3pt}
\noindent
\textbf{Acknowledgements.}
This work is partially supported by a research grant from the National Natural Science Foundation of China under Grant 62306254 and a grant from the Hong Kong Innovation and Technology Fund under Grant ITS/030/21.

\vfill


{
    \small
    \bibliographystyle{ieeenat_fullname}
    \bibliography{main}
}

\end{document}